\documentclass[aps,prc,showpacs,superscriptaddress,preprint,floatfix]{revtex4-1}

\usepackage[dvips]{graphicx}
\usepackage{color}
\usepackage{amssymb}
\usepackage{amsmath}
\usepackage{braket}
\usepackage{hyperref}
\usepackage{bm}
\usepackage{multirow}

\newcommand{\nc}{\newcommand}           
\nc{\vc}[1]     {\mbox{\boldmath $#1$}} 
\nc{\mapleft}[1]{                       
 \smash{\mathop{                        %
  \hbox to 0.90cm{\rightarrowfill} }\limits_{#1}}}
\nc{\figwidth}{0.8}                    

\nc{\mydraft}	{\setlength{\topmargin}{-1.5cm}}
\mydraft   

\begin{document}
\title{Role of unitary correlation operator on high-momentum antisymmetrized molecular dynamics using bare $NN$ interaction for $^3$H and $^4$He}

\author{Qing Zhao} \email[]{zhao@nucl.sci.hokudai.ac.jp}
\affiliation{Nuclear Reaction Data Centre (JCPRG), Hokkaido University, Sapporo 060-0810, Japan}

\author{Masahiro Isaka}
\affiliation{Hosei University, 2-17-1 Fujimi, Chiyoda-ku, Tokyo 102-8160, Japan}

\author{Takayuki Myo}
\affiliation{General Education, Faculty of Engineering, Osaka Institute of Technology, Osaka, Osaka 535-8585, Japan}
\affiliation{Research Center for Nuclear Physics (RCNP), Osaka University, Ibaraki, Osaka 567-0047, Japan}

\author{Mengjiao Lyu}
\affiliation{College of Science, Nanjing University of Aeronautics and Astronautics, Nanjing 210016, China}

\author{Hiroshi Toki}
\affiliation{Research Center for Nuclear Physics (RCNP), Osaka University, Ibaraki, Osaka 567-0047, Japan}

\author{Hisashi Horiuchi}
\affiliation{Research Center for Nuclear Physics (RCNP), Osaka University, Ibaraki, Osaka 567-0047, Japan}

\author{Hiroki Takemoto}
\affiliation{Osaka University of Pharmaceutical Sciences, Takatsuki, Osaka 569-1094, Japan}

\author{Niu Wan}
\affiliation{School of Physics, Nanjing University, Nanjing 210093, China}


\begin{abstract}
We extend the high-momentum antisymmetrized molecular dynamics (HMAMD) by incorporating the short-range part of the unitary correlation operator method (UCOM) as the variational method of finite nuclei. In this HMAMD+UCOM calculation of light nuclei, the HMAMD is mainly in charge of the tensor correlation with up to the four-body correlation, while the short-range correlation is further improved by using the UCOM. The binding energies of the $^3$H and $^4$He nuclei are calculated with this HMAMD+UCOM using the AV8' bare nucleon-nucleon ($NN$) interaction. The different roles of the short-range and tensor correlations from the HMAMD and UCOM are analyzed in the numerical results. Compared with the previous calculations based on the different variational methods, this newly extended HMAMD+UCOM method can almost provide the consistent results with the $ab$ $initio$ results.

\end{abstract}

\maketitle

\section{Introduction}
In the recent decade, the existence of the strong short-range repulsion and tensor interaction in the bare nucleon-nucleon ($NN$) interaction is confirmed to provide strong short-range and tensor correlations by the $ab$ $initio$ studies of many-body physics in nuclei \cite{Pieper2001, Jastrow1955, Wiringa1995}. The short-range correlation contributes dominantly to the very high-momentum region around $4$ fm$^{-1}$, while the tensor correlation is strong in the midium-momentum region around $2$ fm$^{-1}$ \cite{Lyu2020}. Therefore, the description of the $NN$ correlations including high-momentum component becomes more and more essential for a developed framework to perform the $ab$ $initio$ calculation with the bare $NN$ interaction. 

When reproducing the nuclear structures and the binding energies of light mass region, one of the most successful calculations is the Green's function Monte-Carlo method (GFMC), where various correlation functions with many variational parameters are used for the nuclear wave function without any renormalization \cite{Pieper2002, Pudliner1997}. Besides, in recent years, the frameworks of tensor-optimized shell model (TOSM) \cite{Myo2007PTP, Myo2007PRC, Myo2009} and tensor-optimized antisymmetrized molecular dynamics (TOAMD) \cite{Myo2015, Myo2017PLB, Myo2017PRC-1, Myo2017PTEP-1, Myo2017PRC-2, Lyu2018PTEP} are developed successfully to reproduce the binding energies of s-shell nuclei using the AV8' bare interaction. In TOSM, two-particle two-hole excitations induced by the $NN$ interaction are fully included to get the converging energy. In TOAMD, the correlation functions are introduced for the short-range correlation and tensor correlation and their functional forms are determined to minimize the total energy of the nucleus. In both methods, the high-momentum motion of nucleons in nuclei are explicitly treated. The description of the short-range correlation and tensor correlation basing on the microscopic framework remains an essential topic for the understanding of the inter-nucleon correlations.

In addition to the expansion scheme with the correlation functions for nuclei, another framework named the high-momentum antisymmetrized molecular dynamics (HMAMD) \cite{Myo2017PTEP-2, Myo2018, Zhao2019} has been developed for the description of the $NN$ correlation and this method is an extension of the pioneering idea given in Ref. \cite{Itagaki2018}. In HMAMD, the correlated nucleon-pair with high-momentum component is introduced in the nuclear many-body states and this treatment is extendable by increasing the number of correlated pairs at the same time. It is found that the short-range correlation and the tensor correlation are well described within a rather simple form of high-momentum $NN$ pair excitation. Another unique treatment for the correlations between nucleons is the unitary transformation of the uncorrelated state to the correlated many-body states as done by the unitary correlation operator method (UCOM) \cite{Feldmeier1998, Neff2003, Myo2009}. In UCOM, the short-range correlation and the tensor correlation are described with two kinds of unitary correlation operators separately, which have a good property for porting them to the various kinds of nuclear many-body wave functions. In particular, short-range part of UCOM can work nicely to describe the short-range correlation in finite nuclei \cite{Myo2009, Feldmeier1998, Myo2011} and also in nuclear matter \cite{Myo2019, Hu2010}

Thus, it is rather straightforward to combine the HMAMD with UCOM to further improve the description of the inter-nucleon correlations. Under the unitary transformation, the HMAMD still maitains its analytical simplicity while providing a better description of many-body correlations. In our previous work \cite{Myo2009}, we have incorporated the UCOM as well as the S-wave UCOM (S-UCOM) into the TOSM for the calculation of the $^4$He nucleus and obtained better numerical results which are very close to the rigorous calculation. In this work, we will combine the HMAMD with UCOM, namely HMAMD+UCOM, and investigate the applicability of this new variational method by performing the calculations of $^3$H and $^4$He nuclei using AV8' bare interaction. Since the tensor correlation has already been well described within the framework of HMAMD \cite{Myo2017PTEP-2, Lyu2020}, we will adopt only the short-range part of UCOM in the current work. The detailed role of UCOM will be discussed from the numerical results of this new extension. We will also compare these new results with those from previous TOSM+UCOM/S-UCOM calculations as well as the TOAMD calculations.

This paper is organized as follows: Section \ref{sec:waveFunction} explains the framework of the HMAMD wave function and the UCOM transformation, which we are aiming to propose. The numerical results of the HMAMD+UCOM calculations for the $^3$H and $^4$He nuclei are presented and discussed in Sec. \ref{sec:results}, where the comparison of the current HMAMD+UCOM between the TOSM+UCOM/S-UCOM or TOAMD will also be included. The conclusion is given in Sec. \ref{sec:conclusion}.

\section{Framework}
\label{sec:waveFunction}
We shall begin with the many-body Hamiltonian,
\begin{equation}
\hat{H}=\sum_{i} \hat{T}_i - \hat{T}_\text{c.m.} + \sum_{i<j} \hat{V}_{ij} ~,
\end{equation}
where $\hat{T}_i$ and $\hat{T}_\text{c.m.}$ are the kinetic energies of each nucleon and the center of mass, respectively. In this work, we take the bare nucleon-nucleon interaction AV8' \cite{Pudliner1997} for $\hat{V}_{ij}$ as
\begin{equation}
\hat{V}_{ij} = v_{ij}^\text{C}+v_{ij}^\text{T}+v_{ij}^\text{LS}~,
\end{equation}
where $v_{ij}^\text{C}$ denotes the central term, $v_{ij}^\text{T}$ denotes the tensor term and $v_{ij}^\text{LS}$ denotes the spin-orbit interaction. In AV8', the central term $v_{ij}^\text{C}$ includes the short-range repulsive interaction, which describes the short-range correlation when two nucleons are at short distance. Additionally, the tensor interaction $v_{ij}^\text{T}$ is included in the AV8' comparing with the AV4' central interaction. The Coulomb term is not considered in the current work with the same condition as the other works \cite{Myo2009, Myo2015, Myo2017PLB, Myo2017PRC-1, Myo2017PTEP-1, Myo2017PRC-2, Lyu2018} to compare the results with the AV8' interaction.

The wave function we adopt in this work is the HMAMD wave function, which is based on the antisymmetrized molecular dynamics (AMD). For a nuclear system with $A$ nucleons, the AMD basis can be written as
\begin{equation}
\label{eq:amd}
\Phi_\text{AMD} = \mathcal{A}\{\phi_1(\mathbf{r}_1)\phi_2(\mathbf{r}_2)...\phi_A(\mathbf{r}_A)\}~,
\end{equation}
where the single-nucleon wave functions $\phi(\mathbf{r})$ are expressed in the Gaussian wave packet multiplied by the spin-isospin wave function $\chi_{\tau,\sigma}$ as
\begin{equation}
\phi(\mathbf{r},\mathbf{Z}) = (\frac{2\nu}{\pi})^{3/4}\text{exp}[-\nu(\mathbf{r}-\mathbf{Z})^2]\chi_{\tau,\sigma}~.
\end{equation}
Here, the range parameter $\nu$ represents the size of the nucleon and is common for all nucleons in Eq. (\ref{eq:amd}). It will be determined for each nucleus independently to gain the total energy. The centroid position $\mathbf{Z}$ is a complex vector $\mathbf{Z}=\mathbf{R}+i\mathbf{D}$, where the real part $\mathbf{R}$ represents the spatial position of the Gaussian centroid. We put the condition of $\sum_i^A \mathbf{Z}_i=0$ to keep the center of mass at the origin of the coordinates and the total momentum at zero. For the s-shell nuclei $^3$H and $^4$He, the spatial positions of all nucleons are optimized to be $\mathbf{R}=0$ as the result of the energy variation in TOAMD \cite{Myo2017PTEP-1, Myo2017PRC-2}. The imaginary part of $\mathbf{Z}$, which is $\mathbf{D}$, represents the mean momentum of the Gaussian as
\begin{equation}
\braket{\mathbf{p}} = 2\hbar\nu\text{Im}(\mathbf{Z})~.
\end{equation}

\begin{figure}[htbp]
  \centering
  \includegraphics[width=0.8\textwidth]{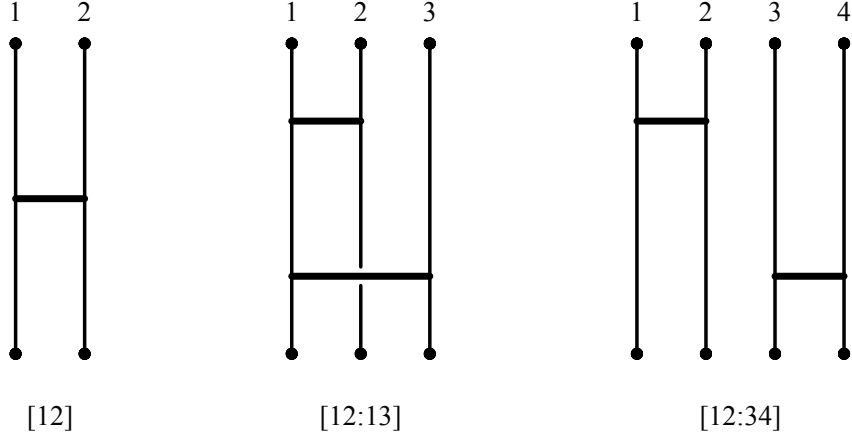}
  \caption{Diagrams of the single and double high-momentum pairs. The vertical lines with the number denote nucleons and the horizontal lines denote the correlations between nucleons. Square brackets below the diagrams represent the configurations of the many-body correlations explained in Refs. \cite{Myo2015, Myo2017PTEP-1}. \label{fig:diagram} }
\end{figure}

In the framework of HMAMD, the imaginary components to the Gaussian centroids of two nucleons $\mathbf{Z}_1$ and $\mathbf{Z}_2$ are assigned with the same magnitude but opposite directions as
\begin{equation}
\mathbf{Z}_1 = i\mathbf{D},~~~~\mathbf{Z}_2 = -i\mathbf{D}~.
\end{equation}
In this way, the high-momentum excitation of di-nucleon pairs is introduced into AMD basis with a large magnitude of $\mathbf{D}$ \cite{Itagaki2018} and also various vector directions \cite{Myo2017PTEP-2}, both of which are necessary to describe the tensor correlation without the center-of-mass excitation. The corresponding AMD basis is extended to the single HMAMD basis $\Phi_\text{1HM}$, since only a single di-nucleon pair is excited. To extend to the double HMAMD basis $\Phi_\text{2HM}$ \cite{Myo2018}, the additional basis with double high-momentum pairs should be included as 
\begin{equation}
\begin{split}
&\mathbf{Z}_1 = i\mathbf{D} \pm i\mathbf{D}^\prime, \\
&\mathbf{Z}_2 = -i\mathbf{D}~,\\
&\mathbf{Z}_3 = \mp i\mathbf{D}^\prime,\\
\end{split}
\end{equation} 
and
\begin{equation}
\label{eq:hmpair}
\begin{split}
&\mathbf{Z}_1 = i\mathbf{D},~~\mathbf{Z}_2 = -i\mathbf{D}~,\\
&\mathbf{Z}_3 = i\mathbf{D}^\prime,~~\mathbf{Z}_4 = -i\mathbf{D}^\prime,\\
\end{split}
\end{equation} 
where up to three or four nucleons are correlated at the same time. The situation in the system with more than four nucleon correlations can be constructed by successively adding more nucleon pairs in Eq. (\ref{eq:hmpair}) in the similar manner, but they are more complicated. The expressions of single and double high-momentum pairs can be explained schematically in Fig. \ref{fig:diagram}.
The left panel shows the single correlation between two nucleons, while the other two panels show the double correlations within three or four nucleons. The double HMAMD basis is required for the precise description of $NN$ correlations in the order of cluster expansion \cite{Lyu2018}.

The spin-isospin parts of the wave function are generally fixed to be $\ket{p\uparrow  n\uparrow  n\downarrow }$ in the $^3$H case and $\ket{p\uparrow p\downarrow n\uparrow n\downarrow}$ in the $^4$He case. However, the spin of the nucleon could be changed due to the tensor operator of $\hat{S}_{12}=3(\sigma_1\cdot {\bf r})(\sigma_2\cdot{\bf r})-(\sigma_1\cdot \sigma_2)$ in the tensor force in the $NN$ interaction. It indicates that the spin parallel configuration of two nucleons can contribute to the total wave function satisfying the Pauli exclusion principle. Therefore, we will add the extra wave functions having the spin-parallel configurations, which is technically treated by using the free spin expression of $\chi_\sigma = \eta\ket{\uparrow}+(1-\eta)\ket{\downarrow}$. Here the parameter $\eta$ is a constant between 0 to 1. The value of $\eta$ will not influence the final result because of the antisymmetrization.

The final wave function of HMAMD is the superposition of all the basis with grid of the momentum shift $\mathbf{D}$:
\begin{equation}
\Psi_\text{HMAMD} = \sum_i f_i \hat{P}^{J^\pi}_{MK}\Phi_\text{1HM}(\mathbf{D}_i)+\sum_{j,k} g_{j,k} \hat{P}^{J^\pi}_{MK}\Phi_\text{2HM}(\mathbf{D}_j,\mathbf{D}^\prime_k)
\end{equation}
where $\hat{P}^{J^\pi}_{MK}$ is the parity and the angular momentum projector. The corresponding coefficients $f_i$ and $g_{j,k}$ will be determined by the diagonalization of the Hamiltonian. For the calculation of $^3$H, $\mathbf{D}_i$ is up to 19.5 fm and meshed by 1.3 fm in the $x$, $y$, $z$ directions. $\mathbf{D}_j$ and $\mathbf{D}^\prime_k$ are up to 13 fm and meshed by 2.6 fm. For the calculation of $^4$He, $\mathbf{D}_i$ is up to 13 fm and meshed by 1 fm. $\mathbf{D}_j$ and $\mathbf{D}^\prime_k$ are up to 10 fm and meshed by 2 fm.

The HMAMD method is a powerful method for the description of the short-range correlation and the tensor correlation, especially for the latter one. On the other hand, another way to implement the correlations into the system is to perform a suitable unitary transformation of the uncorrelated many-body states to a correlated basis, $\widetilde{\Psi} = C\Psi$. Here the tilde notation represents the correlated wave function. By following the works of Feldmeier et al. \cite{Feldmeier1998, Neff2003}, an unitary correlation operator can be written as a product of two unitary operators:
\begin{equation}
C=C_{\Omega}C_r~,~~C_{\Omega}=\mathrm{exp}[-i\sum_{i<j}g_{\Omega,ij}]~,~~C_r=\mathrm{exp}[-i\sum_{i<j}g_{r,ij}]~,
\end{equation}
where $C_\Omega$ is the unitary operator for the tensor correlation and $C_r$ is the unitary operator for the short-range correlation. Two-body operators $g_{\Omega,ij}$ and $g_{r,ij}$ are the Hermite operator where $i$ and $j$ are the particle index. Since the tensor correlation can be well described by the HMAMD basis as studied in Refs. \cite{Myo2017PTEP-2}, we only adopt the short-range correlation part $C=C_r$ in the current work.

In the calculation of UCOM, the Hermite operator $g_{r,ij}$ is defined as
\begin{equation}
g_{r,ij}=\frac{1}{2}\{p_{r,ij}s(r_{ij})+s(r_{ij})p_{r,ij}\}~,
\end{equation}
where the momentum $p_{r,ij}$ is the radial component of the relative momentum,  which is conjugate to the relative coordinate $r_{ij}=|\mathbf{r}_i-\mathbf{r}_j|$. $s(r_{ij})$ is the amount of the shift of the relative wave function of each nucleon pair. In the framework of UCOM, the function $R_+(r)$ can be introduced as
\begin{equation}
\int^{R_+(r)}_r\frac{d\xi}{s(\xi)}=1
\end{equation}
and one can obtain the relation:
\begin{equation}
\frac{dR_+(r)}{dr}=\frac{s(R_+(r))}{s(r)}~.
\end{equation}
The function $R_+(r)$ denotes the transformed relative distance and can reduce the amplitude of the short-range part of the relative wave function in nuclei, which should be determined for four spin-isospin channels (singlet-even, triplet-even, singlet-odd and triplet-odd) independently. Thus, the various operators will be transformed to the form of the relative motion as:
\begin{equation}
\begin{split}
&c^\dagger r c = R_+(r) ~,~~ c^\dagger p_r c = \frac{1}{\sqrt{R'_+}}p_r\frac{1}{\sqrt{R'_+}} ~,~~c^\dagger \mathbf{l} c = \mathbf {l}~,\\
&c^\dagger \mathbf{s} c = \mathbf{s} ~,~~ c^\dagger S_{12} c = S_{12} ~,~~ c^\dagger v(r) c = v(R_+(r))~.
\end{split}
\end{equation}
Here we list the transformation of the relative position operator $r$, the relative momentum operator $p_r$, the relative orbital angular momentum operator $\bf l$, the intrinsic spin operator $\bf s$, the tensor operator $S_{12}$, and the arbitrary scalar function $v(r)$.

The form of $R_+(r)$ depends on the potential adopted in the calculation. For the AV8' potential, the following two parameterizations for the short-range correlation functions, which are proposed by Neff-Feldmeier and Roth et al. \cite{Feldmeier1998, Neff2003, Roth2006, Roth2007}, have proven appropriate:
\begin{equation}
\begin{split}
&R_+^{\mathrm{even}}(r)=r+\alpha\big(\frac{r}{\beta}\big)^\gamma \mathrm{exp}[-\mathrm{exp}(r/\beta)]\\
&R_+^{\mathrm{odd}}(r)=r+\alpha[1-\mathrm{exp}(-r/\gamma)] \mathrm{exp}[-\mathrm{exp}(r/\beta)]~.
\end{split}
\end{equation}
Here, $\alpha$, $\beta$, $\gamma$ are the variational parameters to minimize the energy of the total system, which will be determined for four channels of the spin-isospin pair separately.

With the above unitary correlation operator $C$, one can transform the HMAMD wave function to the correlated one as $\widetilde{\Psi}_\text{HMAMD} = C\Psi_\text{HMAMD}$, and the corresponding Schrodinger equation becomes $\hat{H}\widetilde{\Psi}_\text{HMAMD} = E\widetilde{\Psi}_\text{HMAMD}$ or $\widetilde{\hat{H}}\Psi_\text{HMAMD}=E\Psi_\text{HMAMD}$. Hence, the transformation to the effective Hamiltonian is
\begin{equation}
\widetilde{\hat{H}}=C^\dagger \hat{H} C = \sum_{i} \hat{T}_i - \hat{T}_\text{c.m.} + \sum_{i<j}( \widetilde{v}_{ij}^\text{C}+\widetilde{v}_{ij}^\text{T}+\widetilde{v}_{ij}^\text{LS} + \widetilde{t}_{ij} +\widetilde{t}_{ij}^\text{P}+\widetilde{t}_{ij}^\text{L})~.
\end{equation}
In addition to the correlated nucleon-nucleon interaction term, $\widetilde{t}_{ij}$, $\widetilde{t}_{ij}^\text{P}$, and $\widetilde{t}_{ij}^\text{L}$ are the two-body kinetic terms, which come from the correlated kinetic operator with the cluster expansion by the $\mathrm{C}^\dagger \hat{\mathrm{T}}^{[1]}\mathrm{C}=\hat{\mathrm{T}}^{[1]}+\hat{\mathrm{T}}^{[2]}+\hat{\mathrm{T}}^{[3]}+...$. The number in the square brackets denotes the one-body term, two-body term and so on. For the simplicity of the calculation, we take the approximation to use only up to the two-body operators in the correlated Hamiltonian to treat the short-range correlation, which has been shown to work nicely \cite{Myo2009, Myo2017PRC-1}. In this calculation, we apply the Gaussian expansion technique for the radial part of the effective Hamiltonian, so that the forms of these three kinetic operators are
\begin{equation}
\label{eq:corkin}
\begin{split}
&\widetilde{t}_{ij} = \sum_n c_n e^{a_n r_{ij}^2}\\
&\widetilde{t}_{ij}^\text{P} = \sum_n c^\text{P}_n (\hat{\mathbf{P}}^2e^{a^\text{P}_n r_{ij}^2} + e^{a^\text{P}_n r_{ij}^2}\hat{\mathbf{P}}^2)\\
&\widetilde{t}_{ij}^\text{L} = \sum_n c^\text{L}_n e^{a^\text{L}_n r_{ij}^2}\hat{\mathbf{L}}^2\\
\end{split}
\end{equation} 
where $\widetilde{t}_{ij}$ is a central operator type. $\widetilde{t}_{ij}^\text{P}$ is denoted by using the relative momentum $\hat{\mathbf{P}}$ and $\widetilde{t}_{ij}^\text{L}$ is denoted by using the relative orbital angular momentum $\hat{\mathbf{L}}$. The index $n$ distinguishes the Gaussian range. We also consider the spin-isospin dependence of three operators in Eq. (\ref{eq:corkin}).

In the procedure of the current work, we transform the Hamiltonian to the effective one by the UCOM and then solve the eigenvalue problem of the effective Hamiltonian. We will make the variational calculation for the parameters of the $R_+(r)$ to minimize the total energy results.

\section{Results}
\label{sec:results}
In this work, we combine the HMAMD with the UCOM to calculate $^3$H and $^4$He nuclei using the bare interaction AV8'. Firstly, as a comparison, we show the energy results of $^3$H and $^4$He nuclei only with the HMAMD in Table. \ref{table:energy1}. Since for the HMAMD+UCOM, the optimized size parameters $\nu$ in the single-particle wave function are $0.18$ fm$^{-2}$ and $0.25$ fm$^{-2}$ for $^3$H and $^4$He, respectively, we adopt the same parameters for the HMAMD calculations without UCOM.
\begin{table*}[htbp]
  \begin{center}
    \caption{The energy results of $^3$H and $^4$He nuclei only with the HMAMD. ``1HM" and ``2HM" denote the single HMAMD calculation and double HMAMD calculation, respectively. ``E", ``Kin", ``Cen", ``Ten", and ``LS" denote the total energy, kinetic energy, central term, tensor term and spin-orbit term, respectively. ``Radius" denotes the root-mean-square (r.m.s.) radius with the point nucleon [fm]. All the units of the energy are in MeV. \label{table:energy1}}
    \vspace{2mm}
 \begin{tabular*}{14cm}{ @{\extracolsep{\fill}} l c c c c c c}
    \hline
    \hline
 $^3$H &E &Kin &Cen &Ten &LS &Radius [fm]\\
    \hline
1HM &-5.05   &41.12    &-19.41  &-25.48 &-1.28 &1.68\\
2HM &-6.66   &47.20    &-22.25  &-29.58 &-2.03 &1.61\\
    \hline
 $^4$He & & & & & &\\
    \hline
1HM &-13.07 &77.58 &-40.82 &-48.22 &-1.62 &1.53\\
2HM &-21.01 &89.86 &-51.08 &-57.07 &-2.73 &1.49\\
    \hline
    \hline
  \end{tabular*}
  \end{center}
\end{table*} 
In this table, we see clearly the efficiency of the single and double HMAMD calculations. Adding the double HMAMD basis, the total energies of these two nuclei are surely lowered. The improvement of the total energy comes from both the central part and the tensor part of the Hamiltonian. The kinetic energies also increase in both nuclei because of the high-momentum nature of the tensor correlation. These results indicate that the double HMAMD can contribute to both the short-range correlation and the tensor correlation as introduced in our previous works\cite{Myo2017PTEP-2}. However, the total energies do not reach to the results from GFMC \cite{Kamada2001, Wiringa2002} as $-7.76$ MeV for $^3$H and $-25.93$ MeV for $^4$He. It indicates that even the double HMAMD basis can not fully describe the correlations between nucleons. This is the motivation why we introduce the UCOM upon the HMAMD framework to try to enhance the description of the correlation, in particular, the short-range part, which needs very high-momentum component in the wave function.

Next, we shall investigate the efficiency of the UCOM. We compare the energy results of $^3$H and $^4$He nuclei with single and double HMAMD+UCOM calculations in Table. \ref{table:energy2}. For the results of $^3$H with single and double HMAMD+UCOM calculations, the parameters $\alpha$, $\beta$, $\gamma$ in $R_+(r)$ are optimized in the variational calculation as shown in Table. \ref{table:optpara}. However, for the results of $^4$He, due to the limitation of the numerical calculating power, we skip the variational calculation for the UCOM parameters in the case of double HMAMD calculation and simply use the optimized sets as in the single HMAMD calculation, which is shown in Table. \ref{table:optpara4he}.
\begin{table*}[htbp]
  \begin{center}
    \caption{The energy results of $^3$H and $^4$He nuclei with the HMAMD+UCOM. The meanings of the other symbols are the same as in the previous table. All the units of the energy are in MeV. \label{table:energy2}}
    \vspace{2mm}
 \begin{tabular*}{14cm}{ @{\extracolsep{\fill}} l c c c c c c}
    \hline
    \hline
 $^3$H &E &Kin &Cen &Ten &LS &Radius (fm)\\
    \hline
1HM+UCOM  &-6.27 &46.14 &-22.71 &-28.20 &-1.50 &1.63\\
2HM+UCOM  &-7.08    &48.74   &-23.50  &-30.42 &-1.91 &1.60\\
    \hline
 $^4$He & & & & & &\\
    \hline
1HM+UCOM &-20.32 &90.66 &-54.54 &-54.17 &-2.27 &1.49\\
2HM+UCOM &-23.17 &98.55 &-57.34 &-61.02 &-3.36 &1.46\\
    \hline
    \hline
  \end{tabular*}
  \end{center}
\end{table*} 
\begin{table}[!htbp]
  \begin{center}
    \caption{Optimized parameters in $R_+(r)$ in UCOM for $^3$H calculation. ``1HM" and ``2HM" denote the results of the single HMAMD calculation and the double HMAMD calculation, respectively. \label{table:optpara}}
    \vspace{2mm}
 \begin{tabular*}{10cm}{ @{\extracolsep{\fill}} l c c c c c c c}
    \hline
    \hline
  \multirow{2}*{$^3$H}& \multicolumn{3}{c}{1HM} & & \multicolumn{3}{c}{2HM}\\
  \cline{2-4}\cline{6-8}
  &$\alpha$ &$\beta$ &$\gamma$ & &$\alpha$ &$\beta$ &$\gamma$\\
    \hline
singlet-even &1.30   &0.85   &0.36 & &0.50   &0.75   &0.40\\
triplet-even   &0.50   &0.80   &0.50 & &0.20   &0.70   &1.10\\
singlet-odd   &2.20   &1.26   &1.20 & &1.80   &1.26   &0.70\\
triplet-odd     &1.70   &1.39   &1.00 & &2.10   &1.39   &1.90\\
    \hline
    \hline
  \end{tabular*}
  \end{center}
\end{table} 
\begin{table}[!htbp]
  \begin{center}
    \caption{Optimized parameters in $R_+(r)$ in UCOM for $^4$He calculation. ``1HM" denotes the result of the single HMAMD calculation. \label{table:optpara4he}}
    \vspace{2mm}
 \begin{tabular*}{6cm}{ @{\extracolsep{\fill}} l c c c}
    \hline
    \hline
  \multirow{2}*{$^4$He}& \multicolumn{3}{c}{1HM}\\
  &$\alpha$ &$\beta$ &$\gamma$\\
    \hline
singlet-even &1.30   &0.85   &0.36\\
triplet-even   &0.70   &0.80   &0.40\\
singlet-odd   &1.60   &1.26   &0.70\\
triplet-odd     &1.60   &1.39   &0.90\\
    \hline
    \hline
  \end{tabular*}
  \end{center}
\end{table} 
From Table. \ref{table:energy2}, clear improvements can be found when adopting the UCOM. These facts show that the UCOM can surely provide a better description of the short-range correlation in the framework of HMAMD. When comparing with the results shown in Table. \ref{table:energy1}, we notice that the single HMAMD result of $^4$He provides larger energy gain when adopting the UCOM than in the $^3$H case. It indicates that the UCOM works more effective in the calculation of $^4$He than in the calculation of $^3$H. This difference can be explained from the property of the short-range correlation. Since the short-range correlation describes the strong repulsion when the nucleons are close to each other, it is natural to understand that the compact nuclei such as $^4$He will be more affected by the short-range correlation. This should be the reason why the UCOM for the short-range correlation is more effective in a compact nucleus as $^4$He than in a less compact nucleus as $^3$H.

\begin{figure}[htbp]
  \centering
  \includegraphics[width=0.8\textwidth]{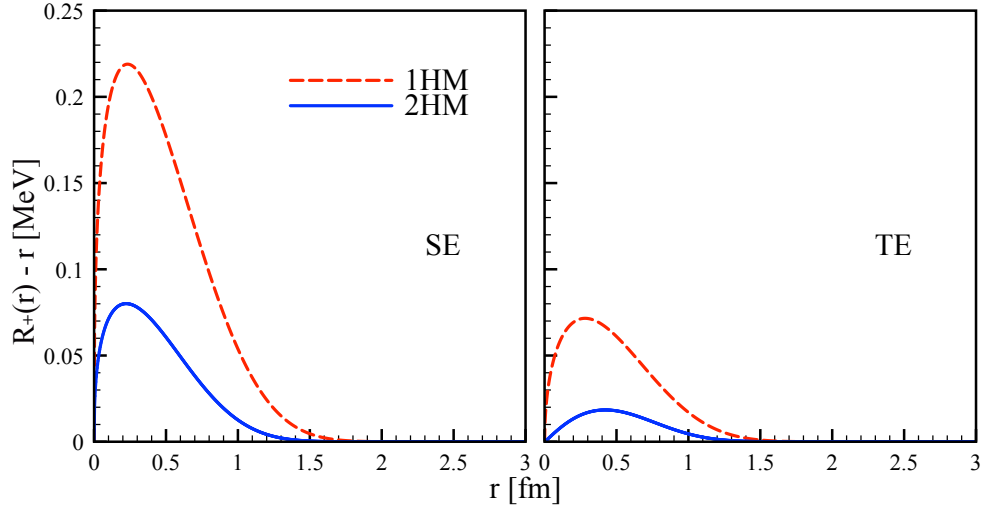}
  \caption{The difference between the shifted distances $R_+(r)$ of the UCOM and the original distance $r$ in singlet-even (SE) and triplet-even (TE) channels. The red dashed lines are optimized for the single HMAMD calculation, while the blue solid lines are optimized for the double HMAMD calculation. Calculations are performed for $^3$H with AV8' potential. \label{fig:optpara} }
\end{figure}

\begin{figure}[htbp]
  \centering
  \includegraphics[width=0.6\textwidth]{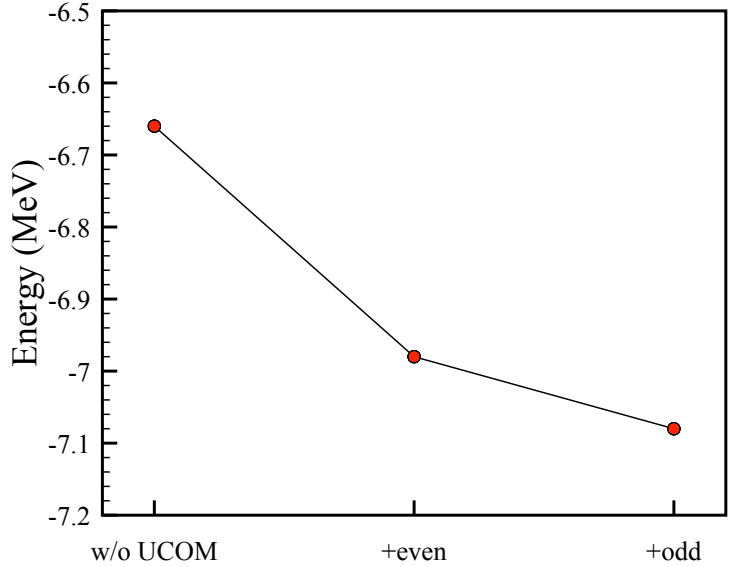}
  \caption{The energy of the $^3$H nucleus calculated with double HMAMD by successively adding the even channels (+even) and odd channels (+odd) of the UCOM. \label{fig:channels} }
\end{figure}

Next, we intend to show how the UCOM effect differs between the single and double HMAMD basis in describing the short-range correlation. The form of $R_+(r)$ for $^3$H calculation are plotted for the even channels in Fig. \ref{fig:optpara}.
From the triplet-even (TE) panel of Fig. \ref{fig:optpara}, it is shown that the strength of the TE channel is small in the calculation comparing with other channels. A reasonable explanation is that the tensor force can also be transformed in the TE central part of the Hamiltonian in the UCOM. However, since the strong tensor effect is already implemented by the tensor part of the AV8' interaction, the strength of the TE channel in UCOM should be reduced to survive the bare tensor force as much as possible. When we compare the UCOM effect with the single and the double HMAMD calculations in the SE and TE channels, we can find that the UCOM effect will be reduced when adding the double HMAMD basis. These results again indicate the well description of the short-range correlation by the double HMAMD basis, so that the UCOM effect becomes smaller than the case of single HMAMD. We also show the energy contributions from the even and odd channels of the UCOM for $^3$H in Fig. \ref{fig:channels}.
In this figure, when introducing only the even channels of the UCOM, the energy will be rapidly reduced while the odd channels contribute a little to the final result. This result shows that the role of the even channel is important rather than the odd channel in the short-range correlation as the discussion reported in Ref. \cite{Lyu2018}.

Next, we add the spin parallel configurations into the calculated model space. This is because the $S_{12}$ tensor operator will change the upward spin parallel states to the downward spin parallel states as $\ket{p\uparrow n\uparrow}\rightarrow \ket{p\downarrow n\downarrow}$, so that the spin parallel configuration of the two neutrons or protons should also be considered. In Table. \ref{table:ffspin}, we show the comparison of the results of $^3$H and $^4$He with the spin parallel or only the natural spin configuration. These calculations are performed with the same parameters for each nucleus and include the double HMAMD basis.
\begin{table*}[htbp]
  \begin{center}
    \caption{The calculation results of $^3$H and $^4$He nuclei with or without the spin parallel configuration. The meanings of the symbols are the same as in the previous table. All the units of the energy are in MeV. \label{table:ffspin}}
    \vspace{2mm}
 \begin{tabular*}{14cm}{ @{\extracolsep{\fill}} l c c c c c c}
    \hline
    \hline
 $^3$H &E &Kin &Cen &Ten &LS &Radius (fm)\\
    \hline
$\ket{p\uparrow n\uparrow n\downarrow}$  &-6.92 &48.18 &-23.44 &-30.01 &-1.66 &1.61\\
$\ket{p\uparrow n\uparrow n\downarrow}+\ket{p\downarrow n\downarrow n\downarrow}$  &-7.08    &48.74   &-23.50  &-30.42 &-1.91 &1.60\\
    \hline
 $^4$He & & & & & &\\
    \hline
$\ket{p\uparrow p\downarrow n\uparrow n\downarrow}$  &-22.10 &95.71 &-56.89 &-58.57 &-2.35 &1.47\\
$\ket{p\uparrow p\downarrow n\uparrow n\downarrow}+\ket{p\downarrow p\downarrow n\downarrow n\downarrow}$ &-23.17 &98.55 &-57.34 &-61.02 &-3.36 &1.46\\
    \hline
    \hline
  \end{tabular*}
  \end{center}
\end{table*} 
From this table, we can observe a slight improvement in the total energy when adding the spin parallel configuration. These results show that the spin parallel configuration is surely required by the bare $NN$ interaction. Furthermore, we can see that the improvement is mainly coming from the tensor interaction, which means that the spin parallel configuration is favored by the tensor correlation as we discussed before. Even though the contribution from the spin parallel configuration is not so large, it can not be ignored especially for the benchmark calculation.

Finally we compare the present results with the TOSM+UCOM/S-UCOM work \cite{Myo2009} and the TOAMD works \cite{Myo2017PTEP-1}. Here, the S-UCOM is the extension of UCOM by introducing the partial-wave dependence in UCOM, where $S$-wave is only transformed in the even channel. In S-UCOM, the tensor contribution increases from the ordinary UCOM because of the enhancement of the $SD$ coupling by the tensor force. In Table. \ref{table:toamd}, we list our current best results for $^3$H and $^4$He as well as the corresponding TOSM+UCOM/S-UCOM and TOAMD results. The $NN$ interaction adopted in these reference works is also the AV8' interaction. The TOAMD method is basing on the AMD basis as is used in our current work, while the shell model wave functions are used in the work with TOSM with the full optimization of the two-particle two-hole excitations. It is also noted that all methods are variational approach using the bare $NN$ interaction directly.
\begin{table*}[htbp]
  \begin{center}
    \caption{The calculation results of $^3$H and $^4$He nuclei from the HMAMD+UCOM, TOSM+UCOM/S-UCOM, and TOAMD. The meanings of the other symbols are the same as in the previous table. All the units of the energy are in MeV. \label{table:toamd}}
    \vspace{2mm}
 \begin{tabular*}{14cm}{ @{\extracolsep{\fill}} l c c c c c c}
    \hline
    \hline
 $^3$H &E &Kin &Cen &Ten &LS &Radius (fm)\\
    \hline
HMAMD+UCOM  &-7.08 &48.74 &-23.50 &-30.42 &-1.91 &1.60\\
TOSM+UCOM     &-6.02 &44.18 &-22.98 &-26.10 &-1.12 &1.70\\
TOAMD \cite{Myo2017PTEP-1} &-7.68 &47.21 &-22.44 &-30.60 &-1.86 &1.746\\
    \hline
 $^4$He & & & & & &\\
    \hline
HMAMD+UCOM  &-23.17 &98.55 &-57.34 &-61.02 &-3.36 &1.46\\
TOSM+UCOM \cite{Myo2009} &-19.46 &88.64 &-56.81 &-50.05 &-1.24 &1.555\\
TOSM+S-UCOM \cite{Myo2009}&-22.30 &90.50 &-55.71 &-54.55 &-2.53 &1.546\\
TOAMD \cite{Myo2017PTEP-1} &-24.74 &97.06 &-53.12 &-64.84 &-3.83 &1.497\\
    \hline
    \hline
  \end{tabular*}
  \end{center}
\end{table*} 
From this comparison, we can first conclude that our HMAMD+UCOM calculation can provide the comparable results as those powerful methods. When comparing with the TOSM+UCOM results, we can find that the HMAMD can gain more energy than TOSM with the UCOM. Furthermore, one can see that the S-UCOM is more efficient than the UCOM with TOSM as shown in the table, which is explained in Ref. \cite{Myo2009}. Thus, it is reasonable to estimate that HMAMD+S-UCOM instead of the UCOM will get lower energy than the present results. The progress may benefit the description of the tensor correlation as it has been done on the TOSM. And when noticing that the TOAMD gains more energy in the tensor part of the Hamiltonian than the HMAMD+UCOM, the S-UCOM could be the solution for the current HMAMD+UCOM work to reproduce or improve the results in comparison with the TOAMD framework. In S-UCOM, we need the partial-wave expansion of the single-particle wave function in AMD \cite{Togashi2009}, which is considered in the future work. The other possible improvement is the addition of the third pair of nucleons having high-momentum in the HMAMD basis states called the triple HMAMD, which is a natural extension of the present double pair case in Eq. (\ref{eq:hmpair}).

\section{Conclusion}
We combine the unitary correlation operator method (UCOM) for short-range part with the double high-momentum antisymmetrized molecular dynamics (HMAMD) as a new variational method for finite nuclei using the bare $NN$ interaction. In HMAMD, up to the two high-momentum pairs excitation is included in the basis states. The UCOM is in charge of the short-range correlation while the HMAMD is expected to be in charge of the tensor correlation. By adopting this new HMAMD+UCOM framework, we calculate the binding energies of the $^3$H and $^4$He nuclei with AV8' interaction, in which the short-range and the tensor correlation are included.

From the results getting from this work, we can conclude that the UCOM has certainly contributed to the energy gain in the final results. We also find that the UCOM has different efficiency in $^3$H and $^4$He cases. It is because the short-range correlation described by the UCOM becomes large in the more compact nuclei. By comparing the calculation results from the single HMAMD and double HMAMD with the UCOM, we find that the double high-momentum pair is essential for the description of both the short-range correlation and the tensor correlation. However, even with the double HMAMD basis, the UCOM is still necessary to improve the description of the short-range correlation. When considering the effect of the tensor operator in the $NN$ interaction, we notice that the spin parallel configuration should contribute to the final wave function. Comparing with the present results from the TOSM+UCOM/S-UCOM and TOAMD, the results getting from this work still have the possibility to be further improved by the S-UCOM, so that has the potential to perform the $ab$ $initio$ calculation in future works.

\label{sec:conclusion}

\begin{acknowledgments}
The authors acknowledge fruitful discussions with Prof. Masaaki Kimura and Prof. Bo Zhou. This work was supported by JSPS KAKENHI Grants No. JP18K03660. Numerical calculations were performed partially in the computer system at RCNP, Osaka University, and the server at theoretical nuclear physics laboratory, Hokkaido University.
\end{acknowledgments}

\bibliographystyle{spphys}
\bibliography{citation}

\end{document}